\documentclass[a4paper,11pt]{article}
\usepackage{pos}
\usepackage{float}

\title{Multi-messenger constraints on transient accelerators of ultra-high energy cosmic rays}

\title{Gamma-rays from Wolf-Rayet stellar winds}

\author*[a]{A. Inventar}
\author[b]{G. Peron}
\author[b]{S. Recchia}
\author[a]{S. Gabici}

\affiliation[a]{Université Paris Cité, CNRS, Astroparticule et Cosmologie \\ 
F-75013 Paris, France}

\affiliation[b]{INAF Osservatorio Astrofisico di Arcetri \\ Largo Enrico Fermi, 5, 50125, Firenze, Italy}

\emailAdd{inventar@apc.in2p3.fr}

\abstract{Gamma-ray observations of young star clusters have recently provided evidence for particle acceleration occurring at stellar wind termination shocks, fueled by the mechanical energy of stellar winds from massive stars. In this work, we explore the possibility that the wind from a single powerful star, whether isolated or part of a cluster, can alone provide sufficient energy to generate gamma-ray emission detectable by current instruments. This scenario is particularly relevant given that a significant fraction of Wolf-Rayet (WR) stars are not found within clusters. To investigate this, we compiled a large sample of WR stars and ranked them based on their wind luminosity divided by the square of their distance, a proxy for their potential gamma-ray flux. We then searched for spatial coincidences between the most promising candidates and cataloged gamma-ray sources. This analysis leads us to propose associations between the stars WR14, WR110, WR111, and WR114 and several unidentified gamma-ray sources. These results suggest that WR stellar winds could represent a distinct and previously unrecognized population of gamma-ray emitters.}

\ConferenceLogo{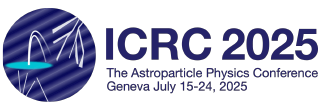}

\FullConference{39th International Cosmic Ray Conference (ICRC2025)\\
 15–24 July 2025\\
Geneva, Switzerland\\}

\begin{document}
\maketitle

\section{Introduction}
\label{sec:intro}

The idea that stellar wind termination shocks (WTS) could accelerate particles and play a role in the observed cosmic-ray (CR) flux has long been considered \citep{casse1980}. Because the mechanical power of stellar winds rises sharply with stellar mass, OB and Wolf-Rayet (WR) stars were quickly identified as plausible sites for such acceleration \citep{cesarsky1983}. In particular, the composition of WR winds, enriched in heavy elements, has been proposed as a possible origin of the excess of $^{22}$Ne observed in CRs \citep{casse1982}. Explaining this enhancement, however, requires the accelerated WR wind material to be significantly diluted within the interstellar medium, implying that the contribution of WTS to the total galactic CR flux is likely modest, though essential \citep{tatischeff2021}.

Recent gamma-ray observations have renewed interest in WTS as acceleration sites, especially following the detection of several young star clusters emitting at high energies \citep{gabici2024}. In these environments, gamma-ray emission may result from interactions between CRs and surrounding matter or radiation fields, either through hadronic processes \citep{aharonian2019} or leptonic mechanisms \citep{harer2023}. However, since star clusters often contain both stellar winds and supernova remnants, it remains difficult to determine which sources dominate the particle acceleration \citep{vieu2022}. An important clue comes from very young star clusters still embedded in their natal molecular clouds, which are too young to have produced supernovae. In these cases, stellar winds are the only viable explanation for particle acceleration up to TeV energies, and the subsequent observed gamma rays. Moreover, the dense surroundings point towards hadronic interactions as the dominant emission mechanism \citep{peron2024a,peron2024b}.

Motivated by these findings, we investigate whether the wind from a single massive star might suffice to generate gamma-ray emission detectable by current instruments, even if the star is located outside of a cluster environment. WR stars are particularly compelling in this context, as they drive some of the most powerful known stellar winds. Although the WR phase is relatively short (on the order of $10^5$ years) the mechanical energy released during this time is comparable to that emitted over the star’s entire main sequence lifetime \citep{crowther2007,seo2018}. This makes WR stars strong candidates for producing observable high-energy signatures linked to wind-driven acceleration.

To explore this possibility, we compiled a sample of WR stars with well-characterized wind parameters and distances, and ranked them by their expected gamma-ray output. For several of the most luminous candidates, we identified spatial coincidences with unidentified gamma-ray sources. The comparison shows that the energy available from the stellar winds is sufficient to account for the gamma-ray emission, supporting the idea that WR stars may constitute a previously overlooked population of gamma-ray sources.

\section{Spatial association with gamma-ray sources}
\label{sec:rank}

The mechanical energy from stellar winds provides the energy reservoir for CR acceleration at the WTS of WR stars. Accurately quantifying this energy is therefore essential. By fitting optical and UV spectra with stellar atmosphere models, \citep{sander2019} and \citep{hamann2019} derived the mass-loss rates ($\dot M $) and wind velocities ($u_w$) for presumably single WR stars in the Galaxy. Distances were obtained from Gaia’s second data release (DR2), using parallax measurements.

With precise distances, $\dot{M}$, $u_w$ and hence the wind power $L_w = (1/2) \dot{M} u_w^2$ can be determined more robustly than ever before.
This dataset includes 108 WR stars, accounting for around 15 \% of the Galactic WR population \citep{rosslowe2015}\footnote{\url{https://pacrowther.staff.shef.ac.uk/WRcat/index.php}}.

\begin{figure}[H]
   \centering
   \includegraphics[width=0.8\hsize]{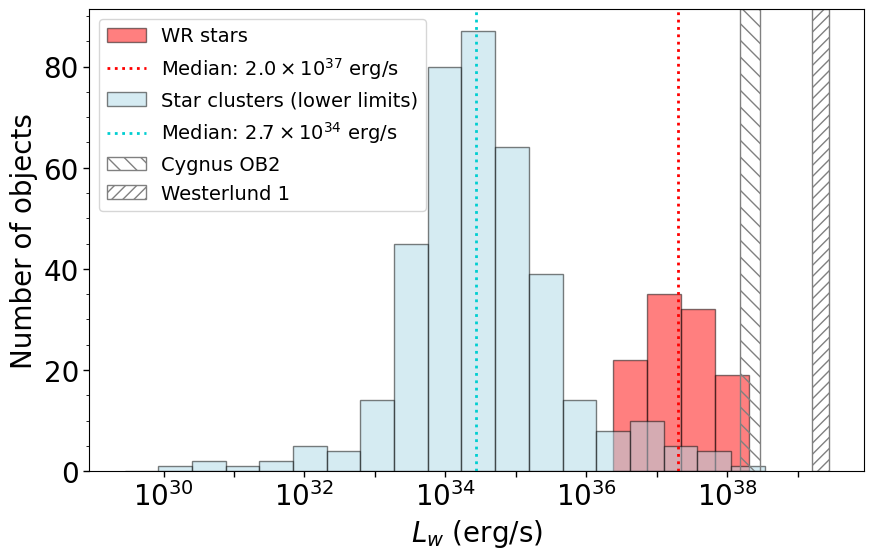}
   \caption{Distribution of wind powers for single WR stars (red) and star clusters (lower limits from \citep{celli2024}, cyan). Estimated wind powers for Cyg OB2 and Wd1 are shown in grey.}
              \label{fig:histo}%
    \end{figure}

The red histogram in Fig.~\ref{fig:histo} shows the resulting distribution of wind powers, which span from a few $\times 10^{36}$ to over $10^{38}$ erg/s, with a median roughly at $2 \times 10^{37}$ erg/s. For comparison, we show lower limits on cluster wind powers from \citep{celli2024}, estimated by scaling Gaia DR2-derived stellar masses to $\dot{M}$ and $u_w$ across cluster members. Because cluster masses are underestimated in this method, the resulting $L_w$ values are conservative. To assess how much, we show in the plot estimates of the global wind mechanical power of Westerlund~1 (Wd1) and Cyg OB2. 

With $L_w \sim 1.6$–$2.8 \times 10^{39}$ erg/s \citep{harer2023}, Wd1's wind power exceeds the lower limit computed in \citep{celli2024} ($>3.15 \times 10^{38}$ erg/s) by a factor of 5–9. This extreme output stems from its rich population of massive stars: 23 WR stars and >160 stars above 25 $M_{\odot}$ \citep{clark2020}. Even increasing cluster powers by a factor 10, individual WR stars still rival the most powerful clusters. Wd1, among the Galaxy's most luminous clusters (with the Nuclear cluster), is detected in gamma rays \citep{aharonian2022}. Similarly, Cygnus OB2 ($L_w = 1.6$–$2.9 \times 10^{38}$ erg/s \citep{menchiari2024}), powered by 3 WR stars and several O stars \citep{vieu2024}, shows gamma-ray emission \citep{ackermann2011,aharonian2019,lhaaso2024,hawc2021} and is only about twice more powerful than the largest WR stars' wind power in our sample.

This motivates a targeted search for gamma-ray emission from single WR stars, especially the isolated ones, which represent over 60 \% of the total population \citep{rate2020}. To identify the best candidates, we rank WR stars by $L_w/d^2$, which scales with the expected gamma-ray flux. 
The ranking is shown in Table~\ref{tab:WRs}.
We consider the WRs with $L_w/d^2 \geq 10^{37}$ erg/s/kpc$^2$, corresponding to the threshold for the faintest gamma-ray-emitting clusters in \citep{aharonian2019}, and this gives a sample of 16 WRs.

\begin{table*}
\begin{center}
\begin{tabular}{| c | c | c | c | c |}
\hline
WR & Cluster/Association & $L_w/10^{37}$ & $d$ & $L_w/d^2/10^{37}$ \\
& & [erg/s] & [kpc] & [erg/s/kpc$^2$] \\
\hline
142 & Berkeley~87 & 12.5 & 1.65 & 4.6 \\
\hline
110 & (Sgr~OB1) & 10.6 & 1.58 & 4.2 \\
\hline
147 & -- & 5.03 & 1.20 & 3.5 \\
\hline
144 & Cyg~OB2 & 9.32 & 1.77 & 3.0 \\
\hline
90 & -- & 3.44 & 1.15 & 2.6 \\
\hline
114 & (Ser~OB1) & 10.0 & 2.10 & 2.3 \\
\hline
52 & -- & 5.86 & 1.75 & 1.9 \\
\hline
15 & (Anon.Vel~b?) & 16.4 & 2.98 & 1.9 \\
\hline
111 & (Sgr~OB1) & 4.18 & 1.66 & 1.5 \\
\hline
136 & (Cyg~OB1) & 5.12 & 1.91 & 1.4 \\
\hline
14 & (Anon.Vel~a:) & 6.22 & 2.23 & 1.3 \\
\hline
25 & Trumpler~16/Car~OB1 & 4.90 & 2.00 & 1.2 \\
\hline
78 & (NGC~6231)/Sco~OB1 & 1.92 & 1.26 & 1.2 \\
\hline
134 & (Cyg~OB3) & 3.65 & 1.74 & 1.2 \\
\hline
6 & Collinder~121 & 5.78 & 2.29 & 1.1 \\
\hline
145 & Cyg~OB2 & 2.13 & 1.46 & 1.0 \\
\hline
\end{tabular}
\caption{WR stars from our sample ranked according to $L_w/d^2$ (wind power/distance squared) and with $L_w/d^2 \ge 10^{37}$~erg/s/kpc$^2$. The star identifier (1$^{\rm st}$ column) is from \cite{rosslowe2015}. The cluster/association membership is also from \cite{rosslowe2015}; "()" indicates that membership is unlikely, ":" that it is likely, "?" that membership was claimed but is now questioned.}
\label{tab:WRs}
\end{center}
\end{table*}

Among these stars, we consider now only the ones that are isolated, using \citep{rosslowe2015}. We then make a cross-match with gamma-ray catalogs, namely the fourth Fermi/LAT catalogue \citep[4FGL, ][]{ballet2023}, 
the H.E.S.S. Galactic Plane Survey (HGPS, \citep{hess2018}),
and the First LHAASO Catalog of Gamma-Ray Sources \citep{cao2024} unveiling respectively the GeV, TeV, and multi-TeV energy domains. To evaluate possible associations with unidentified sources, we start by computing the angular distance ($\Delta \vartheta$) between the centers of the WR stars and of the gamma-ray sources. 
We then select stars that have unidentified gamma-ray sources within an angular distance $\vartheta_s$ corresponding to 30~pc at the distance of the star (see Inventar et al. to be submitted for details). 
Finally, for these stars we search for the minimal values of the quantity $\Delta \vartheta/\vartheta_s$.


\section{Discussion of proposed associations}
\label{sec:unid}

Four matches are found with unidentified sources (using TeVcat \citep{wakely2008} and 4FGL classifications) with this method: WR~114, WR~111, WR~14 and WR~110. 

WR~114 shows no clear association with any stellar cluster or OB association. The star is positioned just 0.06° from the centroid of the Fermi source 4FGL~J1823.3-1340, which has been detected at energies >10 GeV \citep{ajello2017} but lacks a very-high-energy counterpart in TeVCat. The source's 100 MeV-100 GeV flux is $F_{\gamma} \sim 7.5 \times 10^{-11}$ erg cm$^{-2}$ s$^{-1}$, with a log-parabolic spectrum that steepens from $\alpha \sim 2.1$ to $\alpha \sim 2.6$ between 1-10 GeV. Classified as unidentified in 4FGL, the only proposed counterpart in the literature is a tentative association with the globular cluster Mercer~5 \citep{hui2020}.

The spectral steepening favors a hadronic origin for the emission, as a leptonic scenario would require an improbably hard CR electron spectrum ($E^{-3}$ or steeper). We therefore assume that a fraction $\eta$ of WR~114's wind power $L_w$ is converted into CR nuclei (primarily protons). These protons interact with ambient gas of density $n$, producing neutral pions ($pp \rightarrow pp\pi^0$) that decay into gamma rays \citep{kafexhiu2014}. The characteristic timescale for pion production is nearly energy-independent: $\tau_{pp\rightarrow\pi^0} \approx 180 (n/{\rm cm}^{-3})^{-1}$ Myr. Since most of the gamma-ray emission falls within the 100 MeV-100 GeV range, we can express the observed flux as:

\begin{equation}
\label{eq:eq}
F_{\gamma} \approx \frac{\eta L_w}{4\pi d^2} \left( \frac{\tau_{\rm res}}{\tau_{pp\rightarrow\pi^0}} \right)
\end{equation}

where $\tau_{\rm res}$ represents the CR confinement time (limited by the WR lifetime $\tau_{\rm WR}$). For WR~114, with well-constrained $L_w$ and $d$, we derive $\kappa \equiv (\eta/0.1)(\tau_{\rm res}/10^5 {\rm yr})(n/{\rm cm}^{-3}) \sim 7$.\\

WR~111 and WR~14 are associated with slightly more distant and weaker 4FGL sources (J1808.2-2028e with $F_{\gamma} \sim4.5 \times 10^{-11}$ erg cm$^{-2}$ s$^{-1}$ and J0856.0-4724c with $F_{\gamma} \sim1.9 \times 10^{-12}$ erg cm$^{-2}$ s$^{-1}$, respectively). The 'c' flag indicates potential confusion with Galactic diffuse emission, particularly for the latter source whose significance dropped below $4\sigma$ in 4FGL-DR4. We note that WR~111 is also associated to the TeV equivalent of 4FGL~J1808.2-2028e, namely HESS J1808-204. Moreover, as for WR~114, these stars do not show any firm cluster association, and the 4FGL sources with which they are associated have very few counterparts. Their gamma-ray fluxes can be explained with $\kappa \sim 6$ and 0.3, respectively.\\

Finally, WR~110 is associated with the extended TeV source HESS~J1809-193 and lies near the extended Fermi source 4FGL~J1810.3-1925e and the LHAASO source 1LHAASO~J1809-1918u. This complex region contains multiple potential gamma-ray emitters, including pulsars, supernova remnants and molecular clouds. While recent studies \citep{castelletti2016,hess2023,albert2024} have investigated these objects, the origin of the gamma-ray emission remains unclear. The $10^{38}$ erg s$^{-1}$ provided by WR~110's wind could contribute significantly to the observed emission.

\begin{figure*}[ht]
    \centering
    
    \begin{minipage}[t]{0.535\textwidth}
        \centering
        \includegraphics[width=\linewidth]{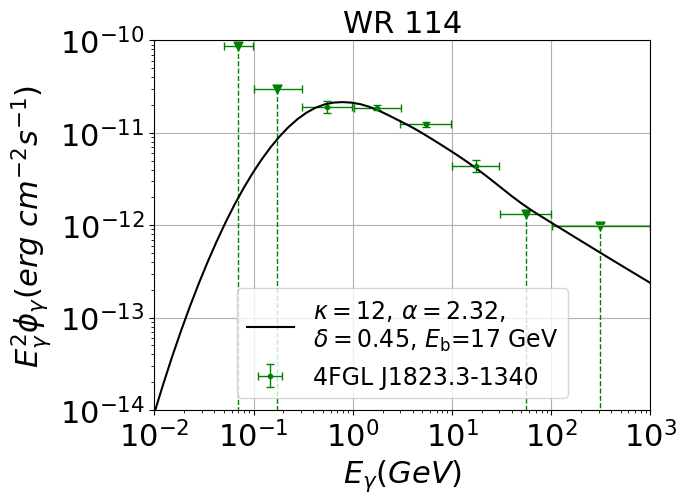}
    \end{minipage}
    \hfill
    \begin{minipage}[t]{0.45\textwidth}
        \centering
        \includegraphics[width=\linewidth]{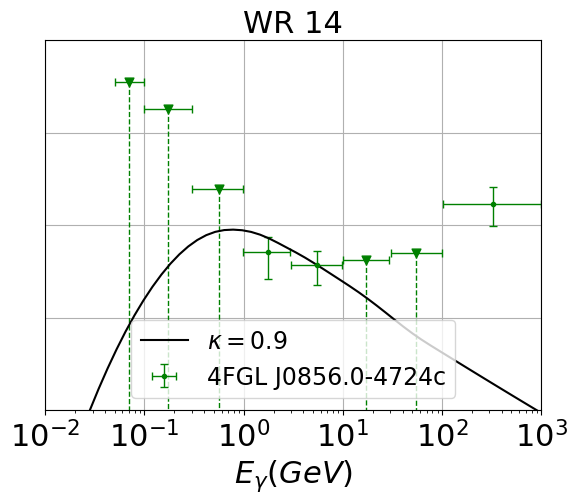}
    \end{minipage}
    \begin{minipage}[t]{0.535\textwidth}
        \centering
        \includegraphics[width=\linewidth]{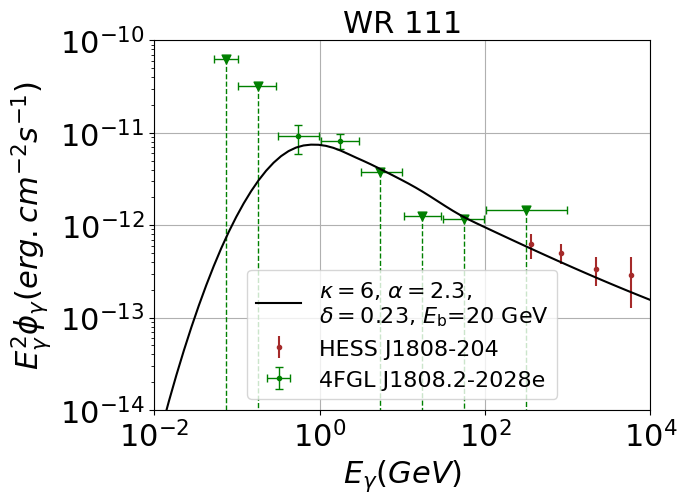}
    \end{minipage}
    \hfill
    \begin{minipage}[t]{0.43\textwidth}
        \centering
        \includegraphics[width=\linewidth]{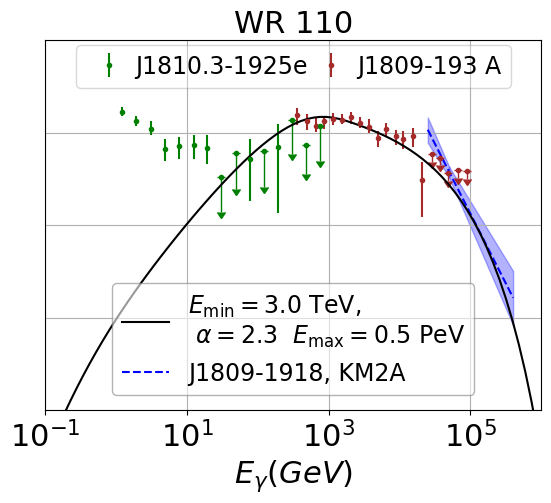}
    \end{minipage}
    \caption{Spectra of the gamma-ray sources associated with WR stars. Solid lines show the predicted $\pi^0$-decay flux (see text). 
    }
    \label{fig:WR_all}
\end{figure*}


Fig.~\ref{fig:WR_all} presents spectral fits for WR~114, WR~111 and WR~110. We also show WR~14's spectra, but the data quality prevents definitive conclusions about its spectral shapes.

We model the WTs of WR~114 and WR~111 as accelerating CR protons with spectrum $Q(E) \propto E^{-\alpha}$, where $\eta L_w = \int {\rm d}E Q(E)E$. Particles above some energy $E_b$ will escape with timescale $\tau_{\rm esc}(E) \sim L^2/D(E) \propto E^{-\delta}$, where $L$ is the system size and $D$ the diffusion coefficient. $E_b$ is then set by $\tau_{\rm esc}(E) = \tau_{WR}$ and the resulting proton spectrum follows $E^{-\alpha}$ below $E_b$ and $E^{-\alpha-\delta}$ above. Gamma-ray emission from $pp$ interactions \citep{kafexhiu2014} scales with our parameter $\kappa$ and best-fit parameters appear in the figure inset. For WR~14, we simply rescale WR~114's prediction by adjusting $\kappa$ and we find values of respectively 12, 6, and 0.9, which is consistent with our earlier estimates from Eq.~\ref{eq:eq}.

For WR~110's complex region, Fig.~\ref{fig:WR_all} shows spectra from HESS~J1809-193 as well as the associated sources 4FGL~J1810.3-1925e and LHAASO~J1809-1918u. The H.E.S.S. collaboration \citep{hess2023} identified two TeV components (A and B) and reanalyzed Fermi data. We show their reanalysis plus H.E.S.S. data for component A ($\Delta \vartheta/\vartheta_s \sim 0.49$). The spectral mismatch between GeV and TeV makes a unified interpretation challenging. Since WR~110 lies near component A and molecular clouds, the TeV emission could come from escaped CRs interacting with dense gas. Modeling this scenario with CR spectrum $\propto E^{-2.3}\exp(-E/E_{\rm max})$ above $E_{\rm min} = 3$ TeV yields a required CR energy of $\sim 6\times10^{47}(n/100 {\rm cm}^{-3})^{-1}$ erg. Compared to the WR's mechanical energy output of $\sim 3\times10^{50}(\tau/10^5 {\rm yr})$ erg, this requires $\sim 0.002(n/100 {\rm cm}^{-3})^{-1}(\tau/10^5 {\rm yr})^{-1}$ conversion efficiency.

\section{Discussion and conclusions}
\label{sec:conclusion}

Our analysis indicates that isolated Galactic WR stars possess sufficient wind power to produce detectable gamma-ray emission with current instruments. To verify this, we conducted a search for spatial correlations between gamma-ray sources and nearby isolated WR stars with strong winds. We identify four promising candidates (WR 114, 111, 14, and 110) where the observed emission can be explained by hadronic processes involving CRs accelerated at the WTS and interacting with surrounding gas.

The predicted gamma-ray flux depends on several factors, with ambient gas density being the most uncertain parameter. After escaping the WTS, CRs initially propagate through the low-density wind cavity before encountering potentially inhomogeneous external material. Consequently, the emission morphology reflects both CR propagation and gas distribution patterns, making our adopted density values effectively averaged quantities.

The search for gamma-ray associations has been made within a radius of 30pc, that roughly corresponds to what is observed for star clusters (see Inventar et al. to be submitted for more details).

While our study examined a substantial WR star sample, it remains incomplete. Future work should incorporate all known WR stars and extend to O-type stars (despite their stronger clustering tendency \citep{maizapellaniz2013}), requiring wind power estimates for a significantly expanded stellar sample.

\end{document}